\documentclass[twocolumn,trackchanges, ]{aastex62}

\usepackage{graphicx,amsmath}
\usepackage{times}
\usepackage[english]{babel}
\usepackage{blindtext}
\usepackage{paralist}
\usepackage{soul}
\usepackage[finalnew]{trackchanges} %for better track changes. finalnew option will compile document with changes incorporated.

\newcommand{\x}{\ensuremath{\times}}

\revised{ApJ}
\shorttitle{AFT+304}
\shortauthors{Upton, Ugarte-Urra, Warren, \& Hathaway}

\begin{document}

\title{The Advective Flux Transport Model: Improving the Far-Side with Active Regions observed by STEREO 304\r{A} }

% \correspondingauthor{}
% \email{}
\author{Lisa A. Upton}
\affiliation{Southwest Research Institute, 1050 Walnut St, Suite 300, Boulder, CO 80302}
\affiliation{Space Systems Research Corporation, Alexandria, VA, 22314}
\author{Ignacio Ugarte-Urra}
\affiliation{Space Science Division, Naval Research Laboratory, Washington, DC 20375, USA}
\author{Harry P. Warren}
\affiliation{Space Science Division, Naval Research Laboratory, Washington, DC 20375, USA}
\author{David H. Hathaway}
\affiliation{Hansen Experimental Physics Lab., Stanford University, HEPL-4085, Stanford, CA 94305-408, USA}

% \collaboration{(AAS Journals Data Scientists collaboration)}

%% --- Abstract
\begin{abstract}
Observations the Sun's photospheric magnetic field are often confined to the Sun-Earth line. Surface flux transport \add{(SFT)} models, such as the {\it Advective Flux Transport} (AFT) model, simulate the evolution of the photospheric magnetic field to produce magnetic maps over the entire surface of the Sun. While these models are able to evolve active regions that transit the near-side of the Sun, new far-side side flux emergence is typically neglected. We demonstrate a new method for creating improved maps of magnetic field over the Sun's entire photosphere using data obtained by the STEREO mission. The STEREO \ion{He}{2} 304\,\AA\ intensity images are used to infer the time, location, and total unsigned magnetic flux of far-side active regions. We have developed and automatic detection algorithm for finding and ingesting new far-side active region emergence into the AFT model. We conduct a series of simulations to investigate the impact of including active region emergence in AFT, both with and without data assimilation of magnetograms. We find that while the \ion{He}{2} 304\,\AA\ can be used to improve surface flux models, but care must taken to mitigate intensity surges from flaring events. We estimate that during Solar Cycle 24 maximum \add[]{(2011-2015)}, $4-6\times10^{22}$ Mx of flux is missing from SFT models that do not include far-side data. We find that while \ion{He}{2} 304\,\AA\ data alone can be used to create synchronic maps of photospheric magnetic field that resemble the observations, it is insufficient to produce a complete picture without direct magnetic observations from magnetographs. 

\end{abstract}


\section{Introduction} \label{sec:intro}

One of the biggest discoveries about the Sun in the last century was \change[]{it's}{its}  magnetic nature. Magnetism on the Sun is ubiquitous and plays a fundamental role in nearly every process occurring on the sun - from the driving the solar cycles \citep{Babcock1961ApJ, Leighton69} to eruptive events like flares \citep{2012LRSP_CME}, which impact space weather and can produce geomagnetic storms here on Earth \citep{baker16}. Unfortunately, magnetic solar observations are currently limited to near-side or Earth-facing side of the Sun. Surface Flux Transport (SFT) models  \citep[see review articles][and references within]{ Sheeley2005, Jiang_etal2014, 2023Yeates_etal} simulate the evolution of the Sun's photospheric magnetic field, creating instantaneous maps of the Sun's magnetic field over the entire surface of the Sun, e.g., 'synchronic maps'. The Advective Flux Transport (AFT) \citep{upton2014a,upton2014b} model is a state-of-the-art SFT model that uses data assimilation of magnetograms combined with a convective simulation to produce synchronic maps with unparalleled realism. AFT has been shown to accurately (within a factor of 2) reproduce the evolution of the total unsigned flux of simple active regions over the course of their lifetimes \citep{ugarte-urra2015}. AFT, and other SFT models, help fill in the blank for the far-side of the Sun, improving our ability to make predictions and space weather forecasts \citep[SFT]{2010Arge_etal,2012MackayYeates, 2015Henney_etal, 2015Schrijver_etal} \citep[AFT]{2021Yardley_etal,2021Warren_etal,2022MackayUpton}.  

While AFT and other SFT models are able to produce maps that include decaying active regions on the far-side, unfortunately new or additional active region emergence that occurs on the far-side is\change{still neglected}{challenging}.  The {\it AFT Baseline} model performs data assimilation of SDO/HMI magnetograms \citep{Scherrer_etal12} to incorporate the near side magnetic flux. The {\it AFT Baseline} maps are the most accurate synchronic maps \change{of the entire Sun}{AFT is currently able to produce with near-side data alone}. However, while these maps continue to evolve the flux that has been observed on the near-side, they still lack new active region emergence that occurs on the far-side of the Sun. To improve upon the {\it AFT Baseline}, we create an new version of AFT which continues to perform data assimilation of SDO/HMI magnetograms, but also includes far-side Active Regions\remove[]{identified in the STEREO/AIA maps}.

\add[]{ 
In the last two decades, several missions to explore and directly observe the far-side of the Sun have been launched, beginning with the twin Solar TErrestrial RElations Observatory} \citep[STEREO]{Kaiser_etal2008} \add[]{satellites in 2006. This pair of satellites entered orbits with one moving progressively ahead of Earth (STEREO A) and one moving progressively behind (STEREO B). Data obtained by pair with the EUVI/STEREO} \citep{Howard_etal2008}  \add[]{instruments enabled stereoscopic imaging of the Sun in the extreme ultraviolet (EUV). By 2011, combining STEREO observations with data obtained by the Atmospheric Imaging Assembly } \citep[AIA]{Lemen_etal2012} \add[]{provided continuous full 360 degree coverage of the active latitudes for nearly four years and at continuous partial far-side coverage for well over a decade. In 2020, a newer mission that directly images the Sun's far-side was launched: ESA/NASA's Solar Orbiter} \citep{Muller_etal2020}. \add[]{ Solar Orbiter's primary mission is not specifically to observe the far-side of the Sun, but rather to fly close to the Sun in an inclined orbit, allowing it to take very high resolution measurements of the Sun and from higher latitudes than can be observed from Earth's orbit. However, its unique trajectory enable it to provide valuable data on the Sun's far-side during portions of its orbit. Most notably, this includes the first far-side magnetograms ever obtained} \citep{Yang_etal2023}. 

In \citet{ugarte-urra2015}, we showed that 304\,\AA\ images can be used as a proxy for magnetic flux measurements when magnetic field data is not available. In this paper, we demonstrate how the 304\,\AA\ proxy can be used to systematically identify far-side active regions observed in 304\,\AA\ and incorporate them into AFT in an autonomous way.  We begin by discussing some revisions that we have made to optimize the flux-luminosity relationship for use with the AFT model. We then describe the process for automatically detecting active regions in the 304\,\AA\ maps and how the far-side active region emergence is incorporated into the AFT simulations. Finally, we discuss the impact of the far-side active region on both active region and solar cycle time scales.

%\textbf{\color{red} Finish writing the Introduction...}

\section{Revised Flux-Luminosity Relationship} \label{sec:relationship}

\begin{figure}[b]
% \centering
% \includegraphics[bb=0 0 340 269, clip, width=1.0\columnwidth]{figures/fluxflux_level2_magtop900_mgthr100_i500_aft_limited.eps}
\includegraphics[width=1.0\columnwidth]{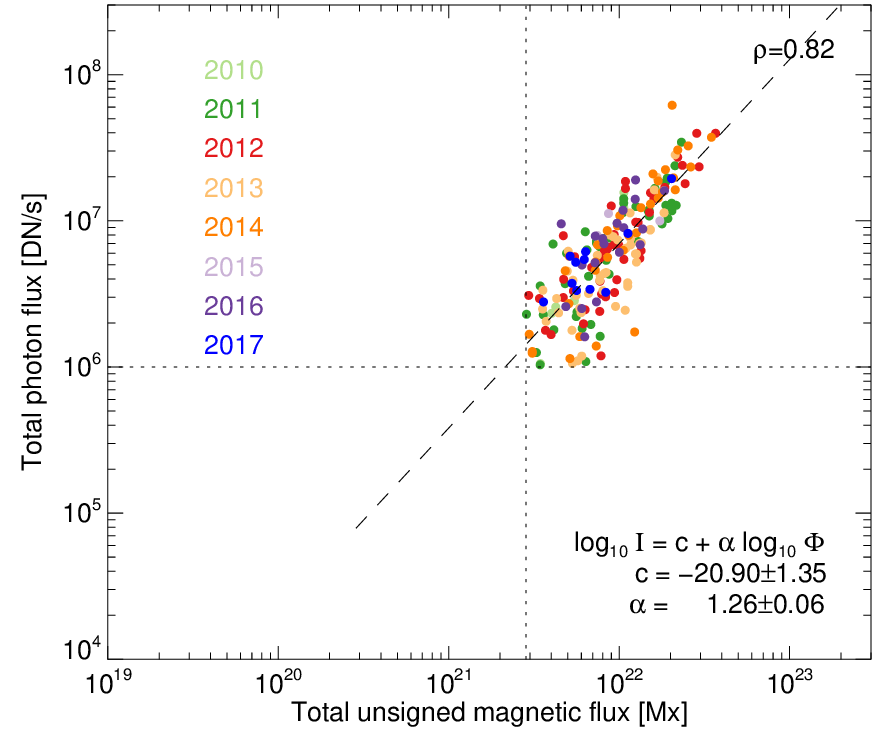}
\caption{Relationship between the 304\,\AA\ photon flux and the total unsigned magnetic flux in the AFT maps of 500 active regions in the period 2010--2018. The photon flux is obtained for full Sun synchronic Carrington maps above a threshold of 500 DN/s/arcsec$^2$. The magnetic flux is calculated for AFT synchronic maps considering magnetic flux densities in the range 100--900G.}
\label{fig:relationship}
\end{figure}

In a previous study \citet[][]{ugarte-urra2015} we showed that \ion{He}{2} 304\,\AA\ images can be used to infer the total unsigned magnetic flux of a region on the Sun when
measurements of the magnetic field are not available. The proxy is given by an empirical power-law
relationship between the total photon flux in 304\,\AA\ and the total unsigned magnetic flux. This method was shown to be sufficient to obtain accurate information about active regions emerging in the far side of the Sun, and was employed to study the continuous long term evolution of active region magnetic flux \citep[][]{ugarte-urra2015,ugarte-urra2017,iglesias2019}.

In the current study we revisit this technique with the objective of systematizing a processing pipeline that consists in: acquisition, calibration and manipulation of 304\,\AA\ images from EUVI/{\it STEREO} and AIA/{\it SDO}, followed by the merger of the images to create a synchronic map of the full Sun that, in its final product, is corrected for center-to-limb variations, sensitivity differences between instruments and sensitivity decay as a function of time. These maps are then used to estimate new empirical relationships for the total photon flux and the total unsigned magnetic flux under various thresholding scenarios. The maps and flux-flux relationships then serve as inputs to the far-side magnetic flux emergence detection algorithm described in Section~\ref{sec:sims}

Several of these steps were already part of the analysis in \citet[][]{ugarte-urra2015}. Here we have introduced a correction for the center-to-limb variation of the
% \ion{He}{ii}
\ion{He}{2} 304\,\AA\ line. We have adopted the parameterization described in \citet[][]{mango1978}: $I(\mu) = I_c [1+a(1-\mu)]$, where $I_c$ is the disk-center intensity, $\mu$ is the cosine of the heliocentric angle, and $a=0.36$ is the relative limb brightening. This factor was obtained from the fits of that curve to solar disk limb-brightening maps made from the median of the weakest on disk intensities in a Carrington rotation. $a$ is the average of the fit coefficient for six Carrington rotations (2055, 2078, 2085, 2095, 2110, 2137). The effect of the center-to-limb variation was also removed in our previous analyses \citep[][]{ugarte-urra2015,ugarte-urra2012}, but was dealt with within a more generic description ("edge effects") and corrected using quiet Sun data from the same map. Before combining AIA and EUVI images into a single map, AIA 304\,\AA\ images are scaled to EUVI intensity  levels with a time dependent factor provided in the SolarSoft STEREO beacon directories \citep[see][]{ugarte-urra2015}. This correction takes care of the sensitivity decay of the AIA instrument. The resulting 304\,\AA\ full Sun synchronic maps (Level 1), with a spatial resolution of 1$\deg$ in longitude and latitude, constitute the main database and are saved with a cadence of 8h, i.e. three files per day.

\begin{figure*}[ht] 
 \centering
 \includegraphics[width=1.0\textwidth]{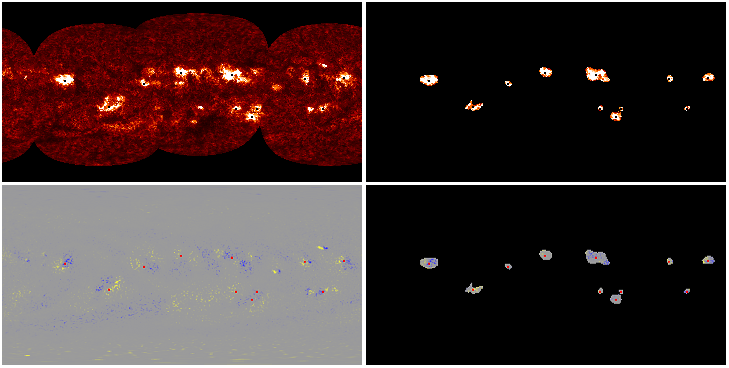}

 \caption{Active Region Detection. The combined STEREO/AIA 304\r{A} map for January 1st 2013 (top left). The active regions identified in the STEREO/AIA 304\,\AA\ map with the automated detection algorithm (top right). The AFT map for January 1st 2013 (bottom left). The AFT active regions that were identified in the STEREO/AIA map (bottom right). The centroid positions of the active regions identified in the 304\,\AA\ image are marked with a black/red squares in the 304\,\AA\ (top right) and AFT (bottom right) images.} 
 \label{fig:detection}
  \end{figure*}

For the purposes of the emergence detection algorithm, we also make a higher order (Level 2) processed file (Figure ~\ref{fig:detection}a). This map is corrected for the temporal decay of EUVI sensitivity, computed from the decay in the centroid value of a histogram of intensities during the years 2007 -- 2019. The centroid for each datapoint per year was taken from the twenty 304 \AA\ synchronic maps in a five day interval at a 6h cadence. The correction we implement is: $f = 1 - 1.268\times10^{-9} \rm dt$ where the corrected intensity is $I_{c}=I_{304}/f$ and $\rm dt$ is the time in seconds since the launch of {\it STEREO} (Oct 25, 2006). This correction amounts to a $f=0.51$ as of January 1, 2019. In these maps we also set to zero regions close to the limb of any of the AIA and EUVI images that make the map, defined as the locations that are beyond 0.9 times the radial distance to the solar disk center. Finally, we remove from the maps intensity spikes, namely flares, with a despiking algorithm available in SolarSoft ({\tt CLEAN\_EXPOSURE.PRO}) designed for cosmic ray despiking in spectra, with the coefficients optimized for our purpose. Pixels identified as spikes are filled in with the median of the surrounding pixels. This step is necessary to avoid the effect that large intensity levels from flares have on the magnetic flux proxy, which relies on 'standard' 304 \AA\ active region intensities.

An example of the corrected STEREO/AIA map for January 1, 2013 is shown in the upper left panel of Figure ~\ref{fig:detection}. We use these new corrected maps to compute new empirical power-law relationships between the total 304 \AA\ photon flux and the total unsigned magnetic flux of active regions. We calculate them from a dataset of 500 active regions observed at central meridian, starting with NOAA 11065 (May 1, 2010) and ending with NOAA 12703 (April 4, 2018), which averages to 1 in 3 NOAA regions in that period. For each region we extract an area of 450$\arcsec$ in Helioprojective Cartesian coordinates ($\sim30\deg$ in Carrington Heliographic Coordinates) and integrate the 304\,\AA\ intensities above a predefined threshold. A threshold of 295 DN/s/arcsec$^2$ (units that the maps are saved in) returns us intensities above 2 times the standard deviation of the map's histogram. In Fig.~\ref{fig:relationship} we show the flux-flux relationship above a threshold of 500 DN/s/arcsec$^2$. The total unsigned magnetic flux in this case is obtained from the {\it AFT Baseline} maps considering only magnetic
flux densities in the range 100 -- 900 G \citep[][]{ugarte-urra2017}. Near central meridian those fluxes are basically the assimilated values from HMI magnetograms. Based on visual inspection of the data, we have discarded data points with less than $1\times10^6$ DN/s and the corresponding magnetic flux given by the linear fit to the power law: $\log_{10} I = C + (\alpha * \log_{10} \Phi)$. This relationship inferred from data observed on the Earth side, allow us to obtain estimates of magnetic flux for regions on the far side for which we do not have magnetic field information.

%{\color{red}[I could add more figures for the various processing steps if needed. Maybe let's focus on the results for now?]}

\section{AFT 304 Automated Detection Algorithm} \label{sec:aft304}

The corrected STEREO/AIA 304 maps are also used to identify far-side active region emergence to input into the AFT model. We have developed an automatic detection algorithm for finding and adding far-side emergence to AFT. The first step of the emergence detection algorithm is to identify the active regions. We begin by creating a binary map of active region locations. We zero out all values in the corrected STEREO/AIA maps where the intensity is less than 500 DN/s/arcsec$^2$. We then smooth the map with a 7 pixel boxcar average filter. All values where the intensity is less than 500 DN/s/arcsec$^2$ are then zeroed out again, producing a binary map of discrete areas that correspond to the active regions. Multiplying the corrected STEREO/AIA 304 map by the binary map clearly reveals the STEREO/AIA 304 active regions and their boundaries. Each discrete active region area is labeled with a unique identifier and a list is generated. 

We then calculate the photon flux (in DN/s) of each active region by multiplying the corrected STEREO/AIA 304 maps by area per pixel and integrating. Using the flux-luminosity relationship derived in the previous section (Figure ~\ref{fig:relationship}), we calculate the proxy magnetic flux ($\Phi_{304}$) for each of the potential active regions. We then exclude regions with a proxy flux less than the threshold of $5.7\times10^{20}$ Mx, the lower limit of the flux-luminosity relationship. The remaining detected active regions are shown in the top right panel of Figure ~\ref{fig:detection}. Finally, we calculate the latitude and longitude of the centroid of each of the remaining active regions (indicated by the square red dots in the image). 

The next step of the automatic detection algorithm is to identify which far-side active regions have new flux emergence ($\Phi_{new}$). This is done by comparing the flux in the STEREO/AIA 304 detected active regions ($\Phi_{304}$) with the flux in active regions in the AFT maps ($\Phi_{AFT}$). We begin by eliminating all near-side active regions (which are presumed to already be included from the data assimilation of HMI maps) by excluding any 304 active region located closer than $\pm$ 75 degrees from the central meridian. For the remaining active regions, we calculate the total unsigned magnetic flux for each active region in the AFT map ($\Phi_{AFT}$), again using the areas from the binary map above (panel d of Figure ~\ref{fig:detection}). We then compare the total unsigned flux for each active region in the AFT map to the proxy flux calculated from the corrected STEREO/AIA 304 maps. If the 304 flux is at least 1.25 times as large as the AFT flux, new flux ($\Phi_{new}$) is assumed to have emerged on the far-side, where 
% \Phi_{new}$ = \Phi_{AFT} - \Phi_{304}.

%
\begin{align}
\Phi_{new} = \Phi_{AFT} - \Phi_{304}.
\end{align} 
\label{eq:fluxvsarea}

\begin{figure}[t]
 \centering
\includegraphics[width=1.0\columnwidth]{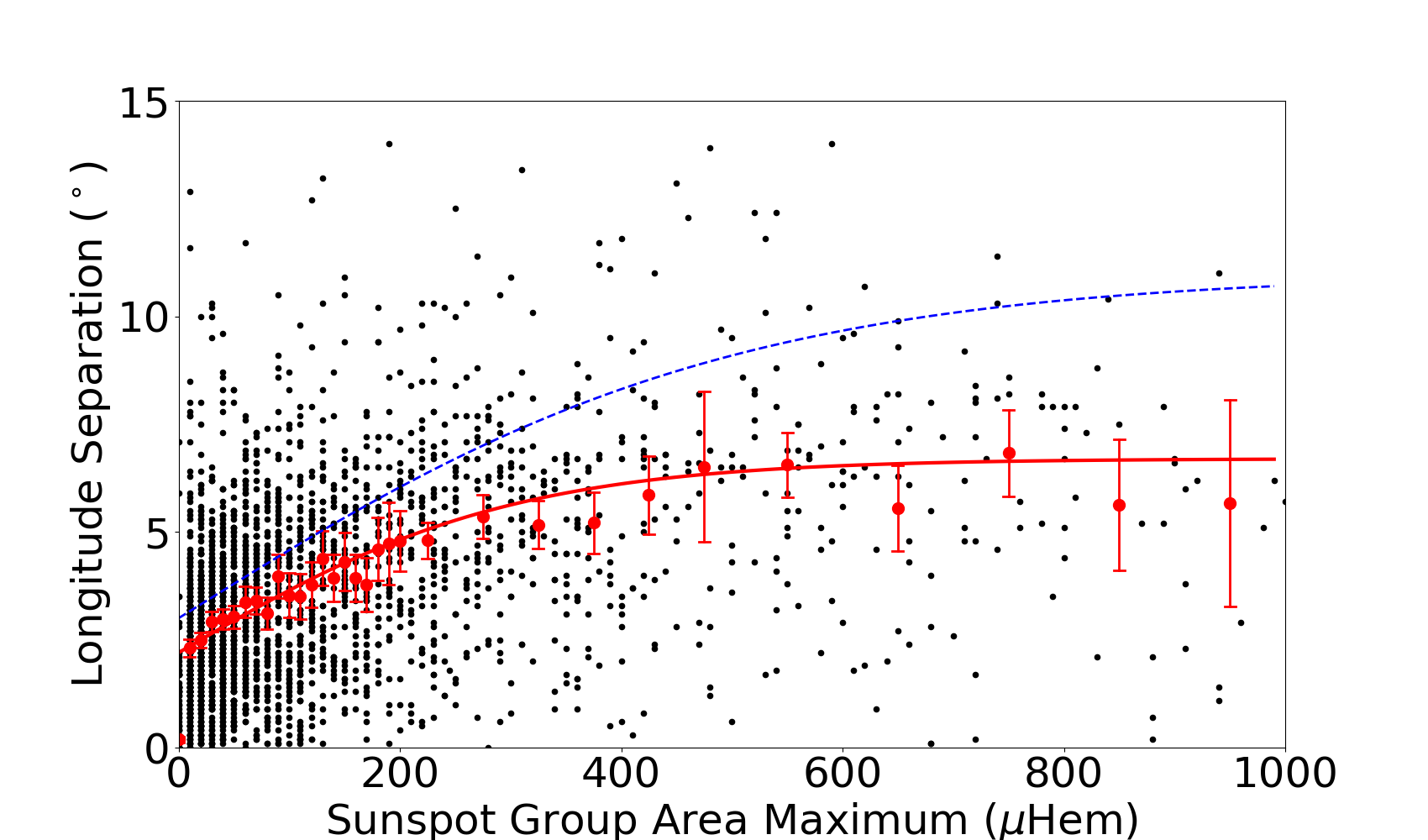}
 \caption{Longitudinal Extent of Active Regions. Black dots represent individual active regions. Red dots represent the average in 31 sunspot area bins with 2 sigma error bars on the averages. The blue line shows the fit used in this work (Equation 3), while the red line shows the recommended fit (Equation 5)}
 \label{fig:longext}
\end{figure}

The final step of the automated algorithm is to add the newly emerged far-side active region flux into the AFT model. This is done by first calculating the area of the active region from flux-area relationship \citep{Sheeley66, Mosher1977}:

\begin{align}
A(\Phi_{304})  &= 7.0\times10^{19}/ \Phi_{304} %\quad[\mbox{Mx}] 
\end{align} 
%\label{eq:fluxvsarea}
%
where $\Phi_{304}$ is the total magnetic flux of the active region in Maxwells as determined by the 304 proxy and $A$ is the total sunspot area in units of micro Hemispheres. Using the total area ($A$) and the latitude of the centroid of the active regions ($\lambda$), we then calculate the bipole tilt angle and bipole separation distance in terms of the latitudinal and longitudinal separations ($\Delta \phi$ and $\Delta \theta$)\change[]{:}{. 

In this work, we use the longitude separations given by} \citet{HathawayUpton2016} \add[]{For a selection of active regions in SOHO/MDI magnetograms, they identified the centroid positions of the bipolar components and calculated the longitudinal separations as a function of the active region area ($\Delta \phi(A)$, where $A$ is in units of millionths of a hemisphere). They found that a good fit to the data (blue line in Figure} \ref{fig:longext} \add[]{) could be found with the following equation:} 

\begin{align}
\Delta \phi(A)  &= 3 + 8 \times \tanh{(A/500)} %\quad[\mbox{Mx}] 
\end{align}
\add[]{ The active region tilts were derived in a similar manner by} \cite{2012StenfloKosovichev}.  \add[]{They found that the tilt as a function of latitude ($\Delta \theta(\lambda)$) could be described by the equation:}   

\begin{align}
\Delta \theta(\lambda)  &= \Delta \phi \times \tan (32.1 \sin (\lambda))
\end{align}

 \remove[]{The longitude separations were determined from magnetic maps of the visible hemisphere obtained with SOHO/MDI.
Centroid positions of the leading and following polarities in each active region were calculated and the relationship given in Eqn. 3 was found to give a good fit. The active region tilts given in Eqn. 4 were derived by} \cite{2012StenfloKosovichev}  \remove[]{using a similar method.} 

We split the new active region flux equally between two opposite polarity bipoles. The bipoles are inserted into AFT as two Gaussian spots with a peak field strength of $\pm$ 900G with Hale's polarity and the latitudinal and longitudinal separations ($\Delta \phi$ and $\Delta \theta$) centered on the centroid of the active region. This is repeated for all active regions where new flux emergence has been identified. AFT is then evolved in time to produce the map for the next time step.
%- Cap flux growth to prevent from adding in too much flux too quickly NOT IN THE BASELINE

\add[]{After the completion of this project, we have conducted a more thorough analysis of longitude separation of the active regions in SOHO/MDI. For 3031 active regions observed by MDI, we identified the day of maximum area for each active region. We then determined the centroid of each polarity in the active region and calculated the longitudinal extent. We divided them into 31 sunspot area bins and found the average (as shown in Figure}  \ref{fig:longext} \add[]{ with 2 sigma error bars on the averages). We find that a better fit (red line in Figure}  \ref{fig:longext} \add[]{) to the longitudinal separations as a function of the active region area ($\Delta \phi(A)$) is given by the following equation: }

%2.5 + 4*tanh(A/300)
\begin{align}
\Delta \phi(A)  &= 2.2 + 4.5 \times \tanh{(A/300)} %\quad[\mbox{Mx}] 
\end{align}
\add[]{We note, that while this equation is a better fit to the MDI observations, more than 80\% of the ARs have maximum area less than 200 and do not expect that the difference in the fit would cause a significant impact to the results presented in this paper. However, we do recommend (and plan ourselves) to use this revised fit to the longitude separation of the active regions moving forward.}

\section{Simulations} \label{sec:sims}

To investigate the impact of including the STEREO/AIA 304 detected active regions, we ran a series of AFT simulations. Each AFT simulation was run from June 1, 2010 through January 1, 2020. In one set of simulations  (referred to as {\it AFT+304}), we performed data assimilation of the HMI magnetograms (as is done in the {\it AFT Baseline}) and we incorporated all STEREO/AIA 304 detected far-side active regions. In the second set of simulations (referred to as {\it AFT 304 only}), ALL the STEREO/AIA 304 detected active regions (near-side and far-side) were incorporated in the AFT model and no data was assimilated from magnetograms. (Note: aside from the initial condition map, these simulation are informed exclusively by the \ion{He}{2} 304\,\AA\ intensity maps.)

\begin{figure}[t]
 \centering
\includegraphics[width=1.0\columnwidth]{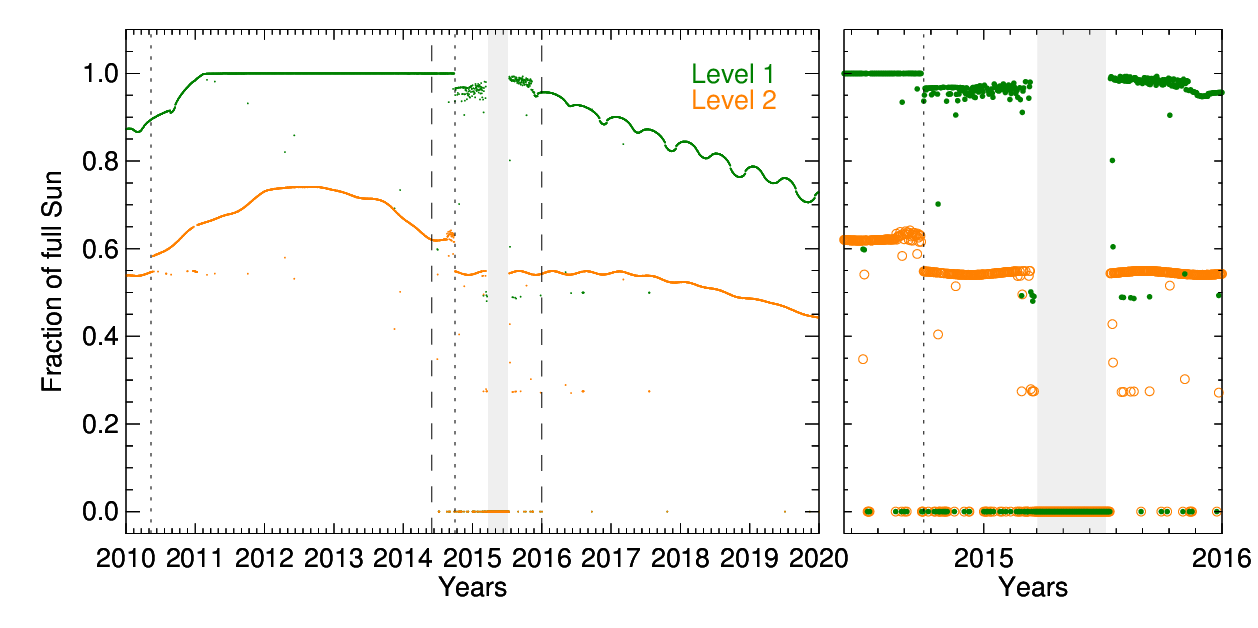}
\caption{Fraction of the Sun observed in the Level1 and Level 2 STERE0/AIA 304 maps as a function of time. The dotted lines highlight when SDO data became available in May 2010 and when communication with STEREO-B spacecraft was lost (October 1, 2014). The shaded area indicates the period of superior solar conjunction when STEREO instruments were turned down. The right panel shows a close-up of the 2014--2016 period delimited by the dashed lines.}
\label{fig:datagaps}
\end{figure}

% \textcolor{red}{STEREO superior solar conjunction mission phase}
 % https://ieeexplore.ieee.org/document/7943731

\begin{figure*}[ht] 
 \centering
 \includegraphics[width=1.0\textwidth]{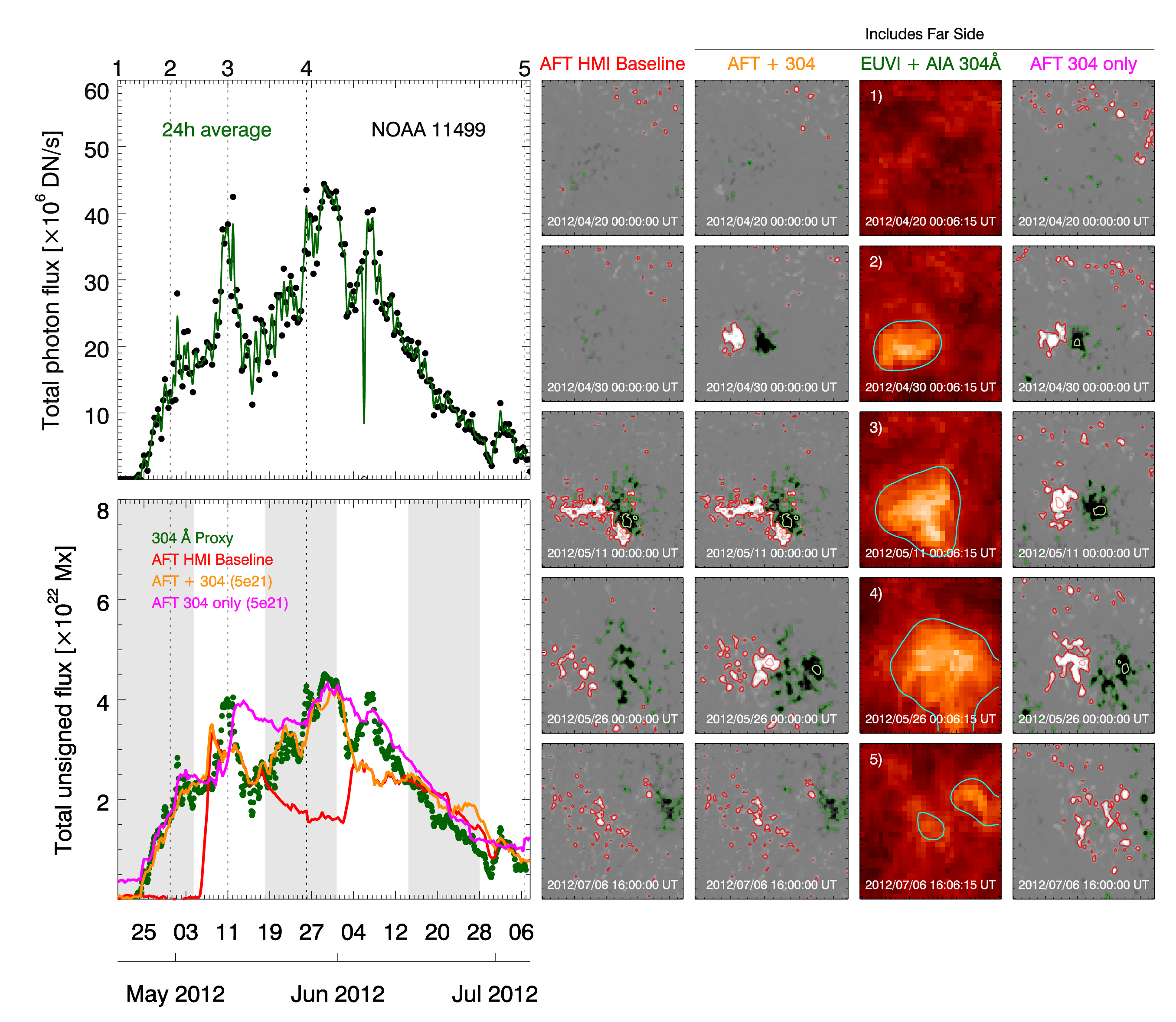}
 \caption{Active Region 11499. We plot the total photon flux in the EUVI/AIA 304 maps for NOAA Active Region 11499 in the top left panel. The bottom left panel shows total unsigned flux as given by the 304 \AA\ proxy (green circles), the {\it AFT Baseline} (red), and {\it AFT+304} (orange). The gray/white background color indicates the availability (not available/available) of nearside magnetic observations. Columns on the right are cutouts of the active region in the {\it AFT Baseline} (left column), {\it AFT+304} (middle column), and EUVI/AIA 304 for select times  during its evolution (marked by the vertical dashed lines in the plots on the left). }
 \label{fig:noaa_11499}
 \end{figure*} 

While we used all available STEREO/AIA 304 Level 2 intensity maps, there are gaps in coverage, beginning the middle of 2014 and lasting through to the end of 2015, that impact the ability to detect far-side active regions during that period. Figure ~\ref{fig:datagaps} illustrates this effect. It shows the fraction of the Sun observed daily in the Level 1 and Level 2 304 maps. This fraction shows a slow secular change due the increase/decrease in far-side area coverage as the  STEREO spacecrafts progress in their orbits, interrupted by two sudden events: a step-function increase in May 2010 when SDO data became available, and a sudden drop in October 2014 when communications with STEREO B spacecraft were lost \citep{cox2018}. In 2014 and 2015 there are many instances where the daily fraction goes to zero due to data loss. The loss of coverage is most notably due to the period (shaded area) of superior solar conjunction when the STEREO spacecrafts were behind the Sun and instruments were powered down, but extends several months, before and after, to a period of limited science operations \citep{ossing2017}. We have also identified several days in 2014 where the gaps in our database do not correlate with gaps in the EUVI database. The latter contribution is less than 6\% of the missing data in that period.

Preliminary results showed that, while the despiking of the STEREO/AIA 304 maps (as described in ~\ref{sec:relationship}) significantly reduced the occurrence of large intensity spikes from flares and false flux deposition, small flares and intensity fluctuations still resulted in spurious flux being added to the AFT simulations. In order to mitigate this, we restricted the amount of flux allowed to emerge per active region per 8h time interval. For each of the two sets of simulations, ({\it AFT+304} and {\it AFT 304 only}), we applied four different threshold caps: (a) $1\times10^{20}$ Mx, (b) $5\times10^{20}$ Mx, (c) $1\times10^{21}$ Mx, and (d) $5\times10^{21}$ Mx, for a total of eight new simulations. %We then compared these six new simulations with the {\it AFT Baseline}. 

\begin{figure*}[ht] 
 \centering
 \includegraphics[width=0.33\textwidth]{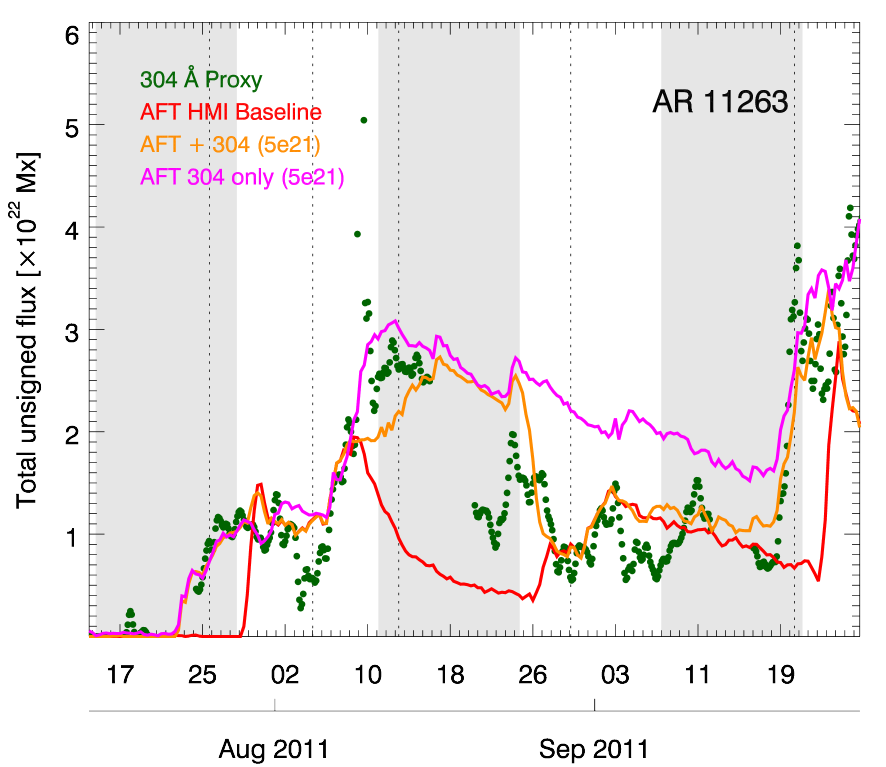}
 \includegraphics[width=0.33\textwidth]{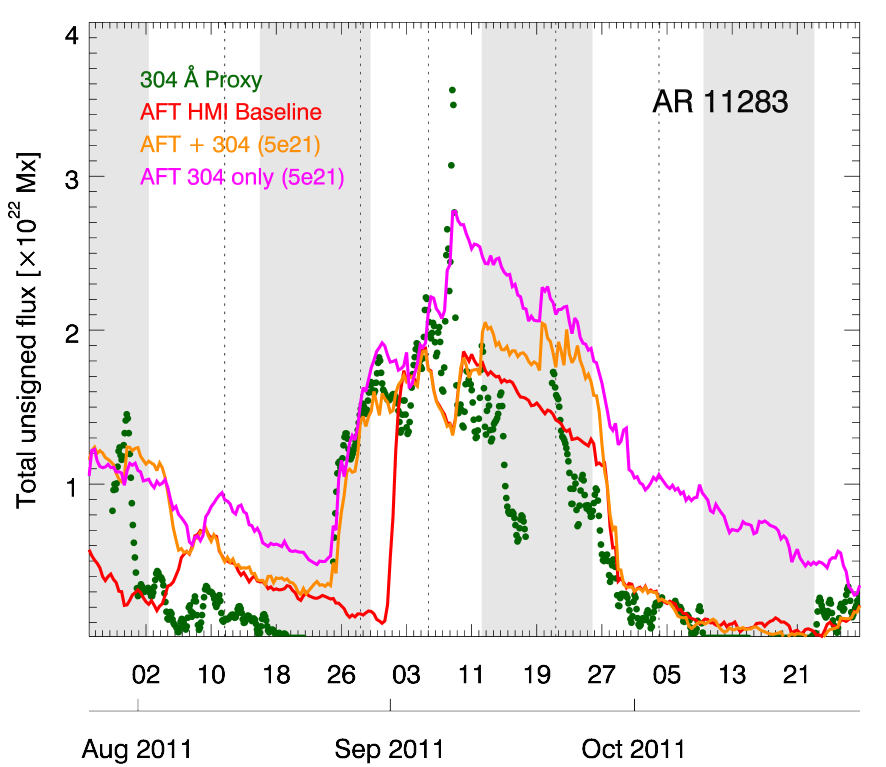}
 \includegraphics[width=0.33\textwidth]{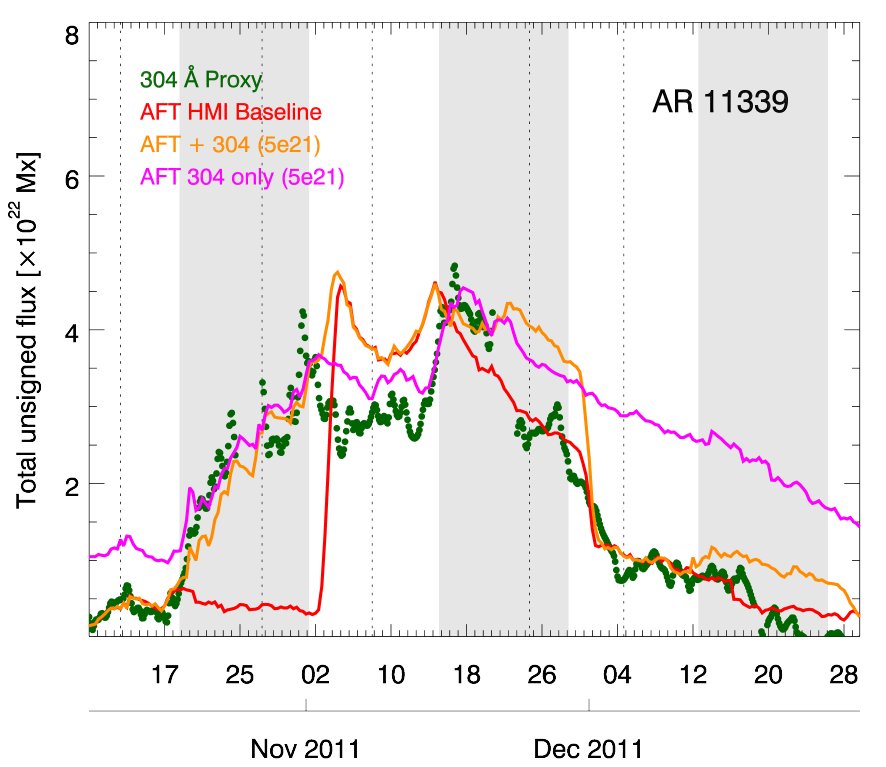}
 \includegraphics[width=0.33\textwidth]{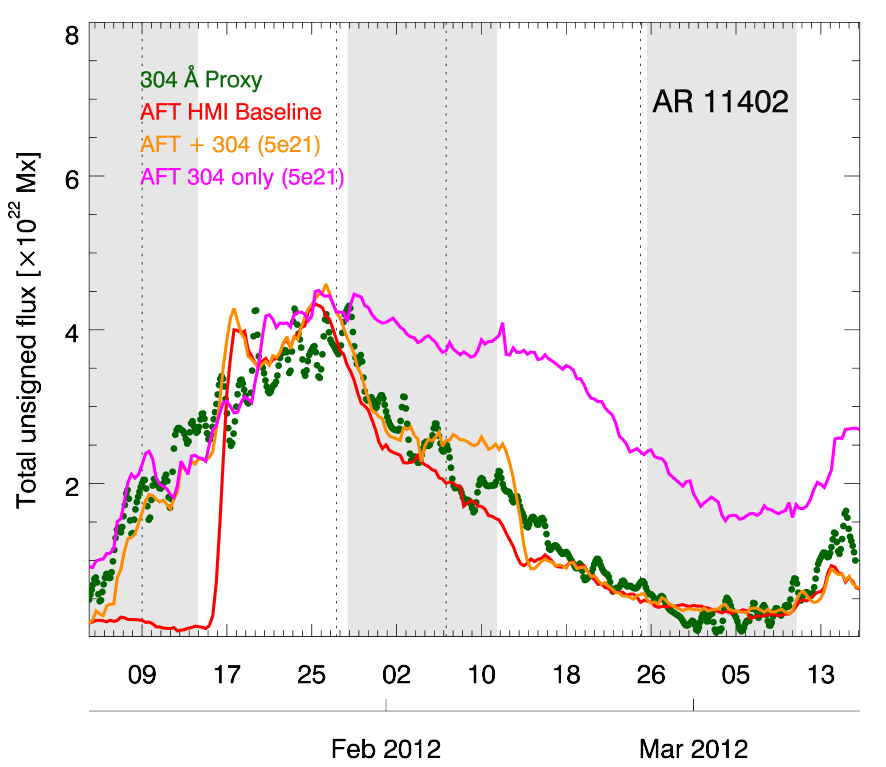}
\includegraphics[width=0.33\textwidth]{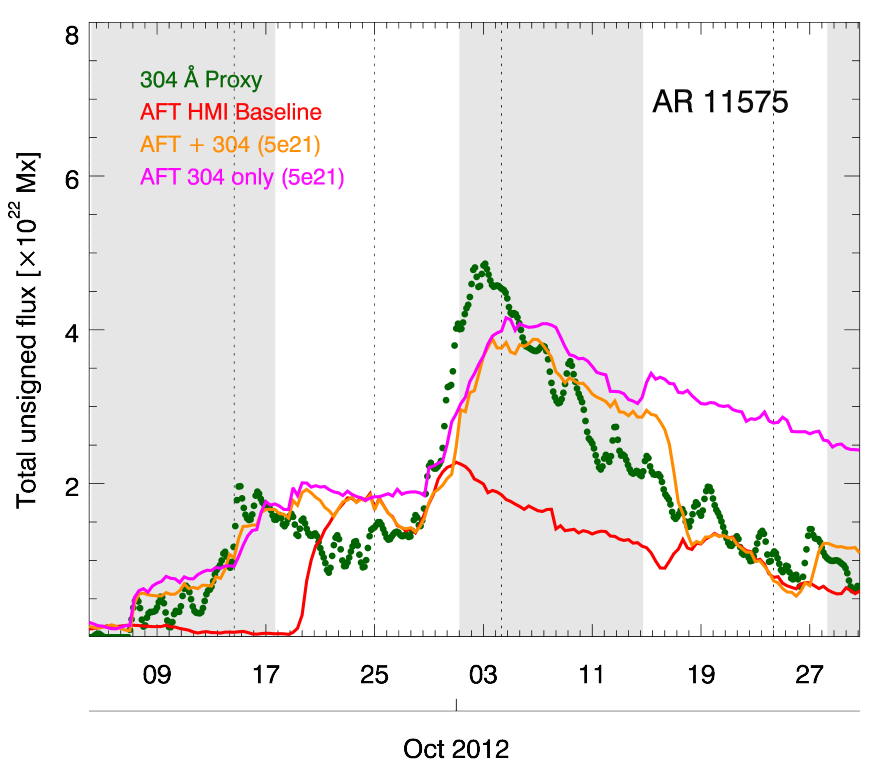}
\includegraphics[width=0.33\textwidth]{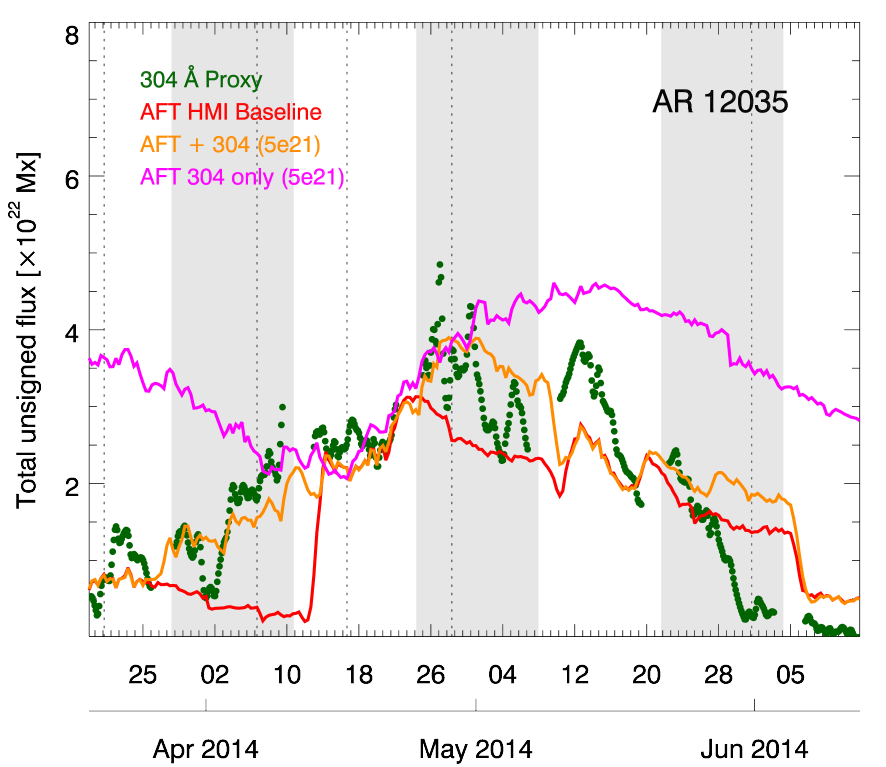}
 %\caption{Active Region Light Curves. \textbf{\color{red}  6 showing comparison of AFT 304 ONLY with 304 Proxy & HMI for the three %thresholds: 1e20, 5e20, 1e21?. Want 6 showing 04 Proxy, HMI, Obs, AFT Baseline, AFt +304, and AFT 304 only (304 with 1e21 threshold). 
% Drop 11499 (in previous figure) and 11968 - only show 11263, 11283, 11339, 11402, 11575, 12035 } }
 \caption{Evolution of the total unsigned magnetic flux for six active regions. Each panel shows the total unsigned magnetic flux estimated from the EUV 304\,\AA\ proxy compared to simulations of the {\it AFT Baseline}, {\it AFT+304} and {\it AFT 304 only} models.} %\textcolor{red}{Fix plot for 11283}}
 \label{fig:AR_compare_AFT304}
  \end{figure*} 

\section{Results} \label{sec:results}

To analyze the impact of using the the STEREO/AIA 304 data in the AFT simulations, we investigate the flux evolution on both small and large temporal and spatial scales. For the small temporal and spatial scales, we examine the evolution of active regions over a few solar rotations. For large temporal and spatial scales, we examine the evolution of global solar properties over several years.

\subsection{Active Regions} \label{subsec:ARs}

% \textbf{\color{cyan} [Begin with AR images from Figure 4 - shows the evolution of an AR over it's lifetime in all simulations.
% Farside captures emergence before the AR appers on the disk
% AFT+304 corrects for shape on near side
% Decay is reasonable well produced in AFT 304 only, but does have some excess flux.
% Demonstrates that this technique is improvement over magnetograms alone]}

% \textbf{\color{red} Best case ARs $-->$ 5e21 }
% \textbf{\color{cyan} Need to show curves that illustrate why 5e21 is best for capturing AR evolution. 
% Discuss the fact that}

% \textbf{\color{cyan} -- 5e21 most effectively captures growth phase. But some flaring still gets though, causing AR flux to be overestimated. This is mitigate by the data assimilation process in the AFT+304.}

%   colors=['medium violet red','orchid','medium orchid', 'magenta']
%   fluxcaps=[1e20,5e20,1e21,5e21]

We investigated the detailed evolution of 7 active regions selected during the period 2011 -- 2014 where we have the best data coverage. The regions (11263, 11283, 11339, 11402, 11499, 11575, 12035) were selected for experiencing flux emergence on the far-side and not too complex in their evolution, although some of them do exhibit multiple emergence episodes during their lifetime (e.g. 11499, 11575). Figure~\ref{fig:noaa_11499} shows the sample case of 11499 as a summary of the analysis performed on each region. First, we computed the total photon flux above threshold (500 DN/s/arcsec$^2$) in the Level 2 304\AA\ maps as the region rotates around the Sun. The top left panel of the figure shows the flux changes observed in active region 11499, with a characteristic intensity increase corresponding to the active region emergence (April 24, 2012), followed by an intensity decrease during the active region decay phase. Interspersed in this secular trend, there are short term variations due to localized additional flux emergence, as well as flares (May 9 -- 11). The 24h averaged photon flux is then used to estimate the total unsigned magnetic flux using the Figure~\ref{fig:relationship} proxy. The resulting flux is plotted in the bottom left panel with green circles. The gray shaded areas in that panel represent the time the region is in transition over the far-side of the Sun.

At this moment, it is worth stopping to compare the magnetic flux proxy curve to the curve returned by the {\it AFT Baseline model} (red line) with no far-side information. The {\it AFT Baseline} model remains oblivious of the presence of active region 11499 until it rotates over the East limb and it is assimilated into the model via the HMI magnetograms. This occurs on May 5, 11 days after the region emerged on the far-side on April 24. Images for key moments for that evolution for the 304 \AA\ maps and the model are shown on the first and third image columns. On April 30, 2012, the region is clearly visible in the EUV 304 \AA\ image, but it is still absent in the {\it AFT Baseline} model. 
Note that we show here the magnetic flux curves with an emergence threshold cap of $5\times10^{21}$ Mx per time step. The discussion of the results obtained using different caps is provided at the end of this section.

\begin{figure*}[ht] 
 \centering
 \includegraphics[width=0.33\textwidth]{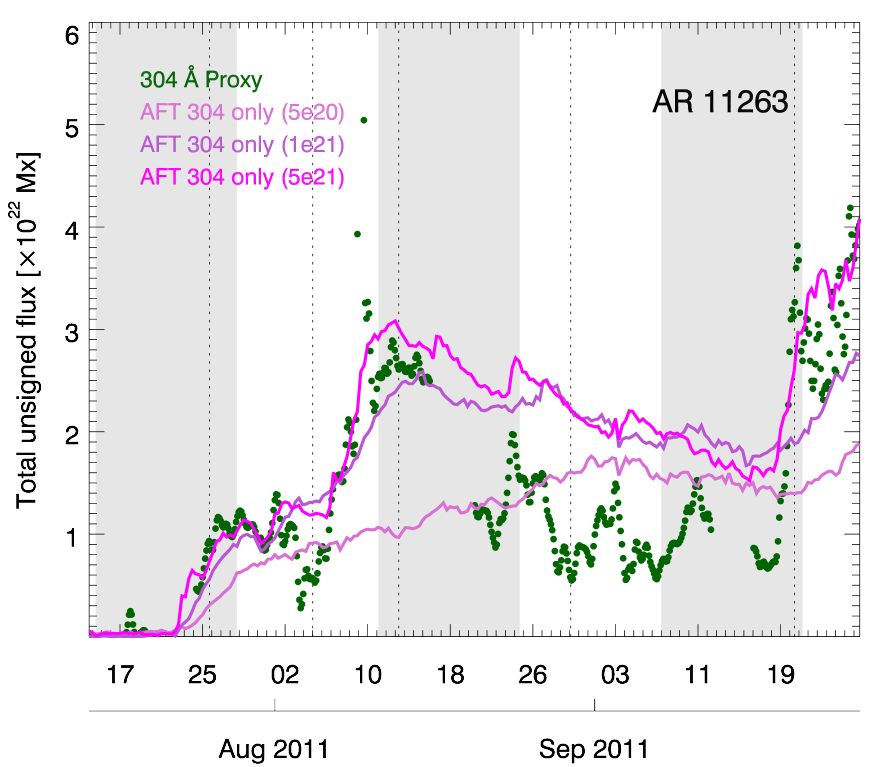}
 \includegraphics[width=0.33\textwidth]{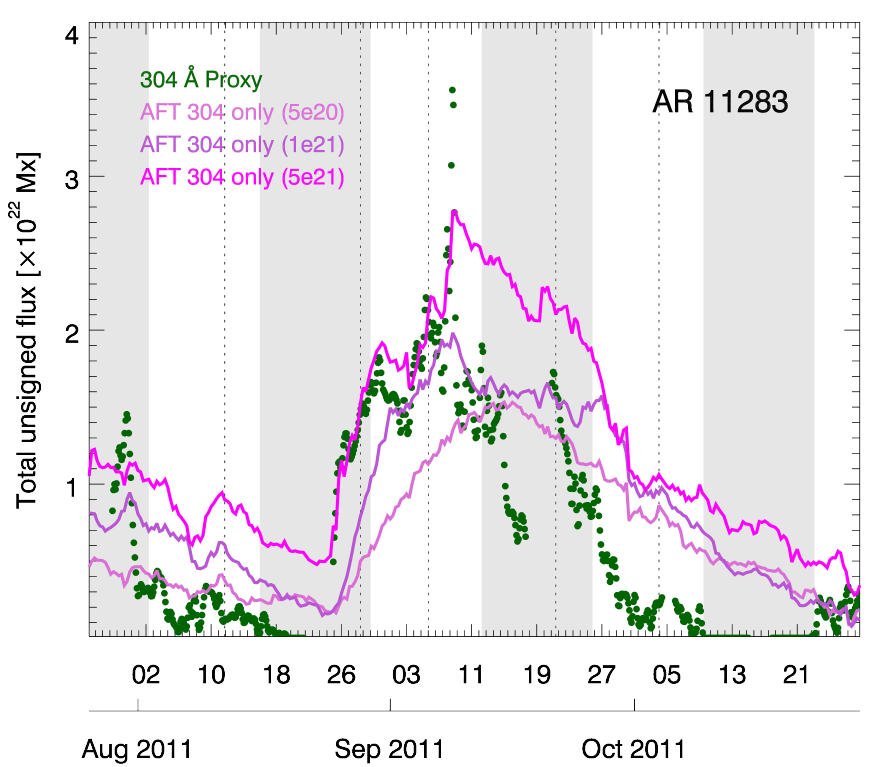}
 \includegraphics[width=0.33\textwidth]{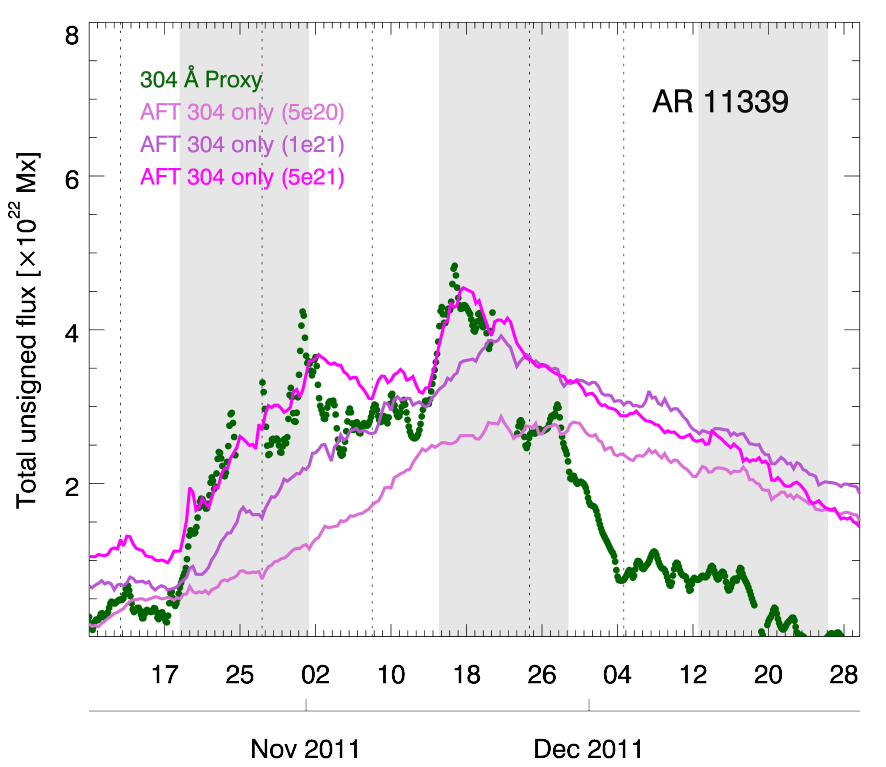}
 \includegraphics[width=0.33\textwidth]{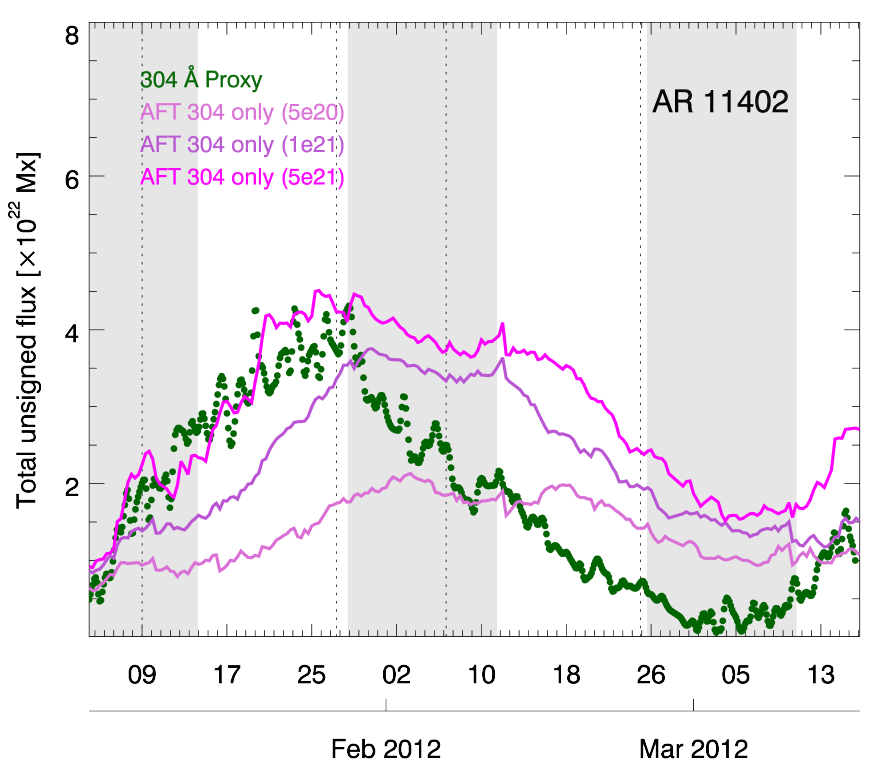}
 \includegraphics[width=0.33\textwidth]{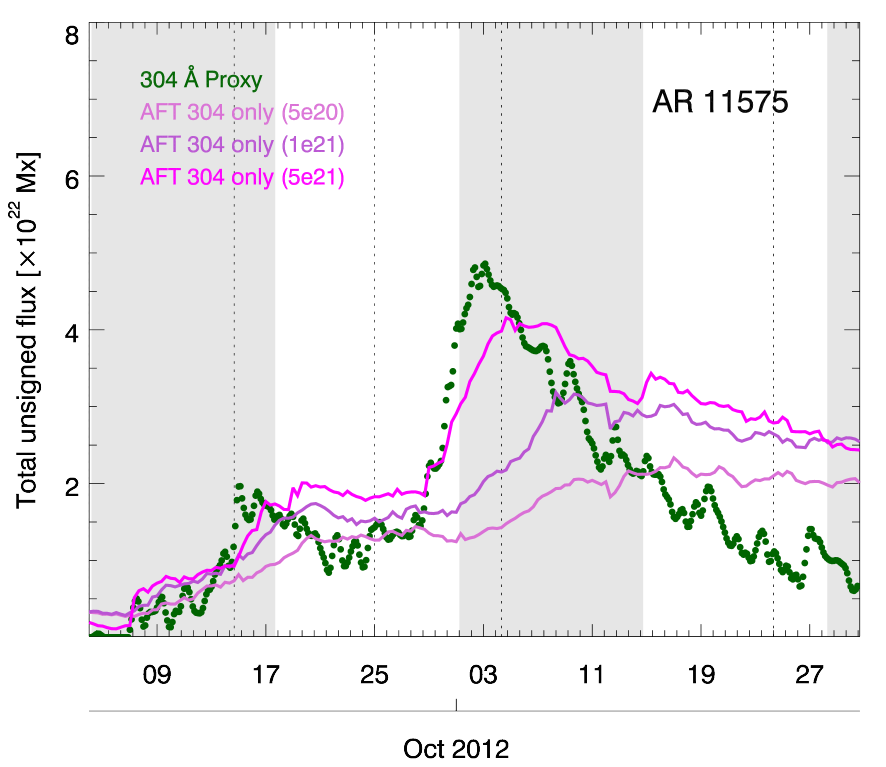}
 \includegraphics[width=0.33\textwidth]{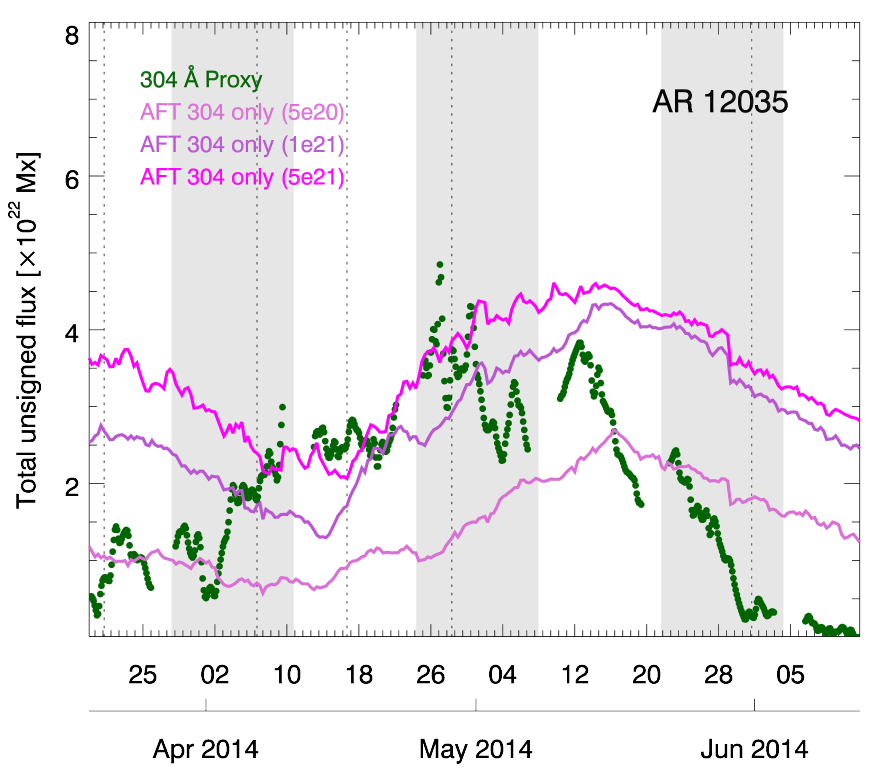}
 \caption{Evolution of the total unsigned magnetic flux for six active regions. Each panel shows the total unsigned magnetic flux estimated from the EUV 304\,\AA\ proxy compared to three simulations of the {\it AFT 304 only} model with three different magnetic flux caps: $1\times10^{20}$, $5\times10^{20}$, $1\times10^{21}$ Mx.}% \textcolor{red}{Fix plot for 11283}}
 \label{fig:AR_compare_304only}
  \end{figure*} 

%   colors=['medium violet red','orchid','medium orchid', 'magenta']
%   fluxcaps=[1e20,5e20,1e21,5e21]

The new {\it AFT+304} simulation, however, which incorporates the far-side emergences from the EUV proxy, does have a bipole emerged at the location of the EUV 304\,\AA\ active region intensity enhancement (Figure~\ref{fig:noaa_11499}, second image column). The corresponding orange magnetic flux curve of the {\it AFT+304} model, seen on the lower left panel, is in good agreement with the magnetic flux proxy discussed earlier. This is an important result because it demonstrates that it is possible to inform a magnetic flux transport model with an EUV imaging proxy to incorporate new far-side flux improving the local magnetic flux fidelity of a purely magnetic model with only near-side information. While both models track each other well while the region is on the Earth side, further emergence on its second far-side pass makes the models diverge again until near-side assimilation makes them converge. In the third far-side pass, the active region does not experience additional flux emergence and simply decays, hence performance of the {\it Baseline} and {\it AFT+304} models is comparable and in agreement with the proxy. 

To understand the validity of the 304\,\AA\ proxy model, we also studied the evolution of the active region in the {\it AFT 304 only} model, where no near-side data is assimilated. This is shown with the pink curve in the bottom left panel and the fourth image column in Figure~\ref{fig:noaa_11499}. The {\it AFT 304 only} simulation is able to incorporate successfully far-side flux emergence using the EUV proxy, improving on the {\it Baseline}. On the Earth side, the {\it AFT 304 only} does not self-correct with data and can diverge from the {\it AFT+304}. Any errors in incorporating short-term spurious new flux, arising for instance from flare emission (e.g. May 11), may result in  unrealistic accumulation of magnetic flux. More along these lines will be discussed in Section~\ref{subsec:global} when looking at the global field. A similar self-correction occurs with the active region tilt, which in the case of the {\it AFT 304 only} simulation follows an estimate from Joy's law.

% SHOW curves like previous paper
% Detailed figure shown for AR 11499
% Additional AR lightcurves to highlight: 11263, 11283, 11339, 11402, 11575,12035
%Not showing 11968(??)

Below we provide a few comments on each of the other six Active Regions investigated in this study: 
\begin{description}
  \item[11263] Active region 11402, below, provides a good example of a simple emergence and decay evolution. Active Region 11263 shows that the model is also capable of dealing with more complex scenarios. The {\it AFT+304} simulation is able to capture 11263 initial far-side emergence, as well as the far-side emergence of a bipolar region to its SE around August 22-24, 2011 (AR 11279), and a large region on September 19 (AR 11302). While not evident from the integrated total unsigned flux curve, the model was able to resolve the different longitudes and latitudes for these three emerging areas within the area of integration.
  
  \item[11283] In the evolution of this region, it is worth highlighting the impact that flare emission can have in the magnetic flux proxy and how the magnetic flux cap mitigates that effect. During the near-side pass, flare activity (September 8-9, 2011) that survived our cleanup process, produces a magnetic flux spike. This spike is ignored by the {\it AFT Baseline} and {\it AFT+304} simulations due to the assimilation process, but it has an impact on the {\it AFT 304 only} case, mitigated by the magnetic flux cap that only allows to add flux at a maximum rate of $5\times10^{21}$ Mx per time step. As the flare intensity decays rapidly, the simulation only incorporates a fraction of the peak flux.
  \item[11339] As in several other examples, the {\it AFT+304} and {\it 304 only} models capture the emergence process of this region in the far side as it happens, whereas the {\it Baseline} model plays catch up with near side assimilated data 15 days later. The 304\,\AA\ proxy underestimates, however, the magnetic flux observed by HMI on the near side. There is an intrinsic uncertainty in the proxy prediction, evident from the spread in Fig.~\ref{fig:relationship}, something that we already discussed in \citet{ugarte-urra2015}. Active region 11339 is a region with a very large sunspot area. Sunspots, while providing significant magnetic flux to an active region, do not correlate well with EUV intensity. While we mitigate this impact in our analysis by setting a magnetic flux integration range of 100 -- 900 G \citep[][]{ugarte-urra2017}, this shows that a simple flux-flux relationship cannot capture the intricacies of all active regions. 
  \item[11402] The simple evolution of this region is a good representation of the ability of the {\it Baseline} model to capture the overall flux evolution in an active region, and how the {\it AFT+304} upgrade extends that to the emergence time. It also reveals the limitations of the {\it 304 only} model. In this case, the discrepancy in the spatial flux distribution compared to the observed one plays a critical role in the cancellation process leading to the decay (i.e. more concentrated in the {\it AFT only} model, more fragmented in the observations). A limitation that the near-side assimilation corrects in the {\it AFT+304} case.
  \item[11575] In the development of this active region, it is interesting to point at the flux emergence taking place on the near-side around September 29, 2012. It is captured both by the assimilated data in the {\it AFT Baseline} and {\it AFT+304}, but also the {\it AFT 304 only} using the proxy. As the region reaches the West limb, the {\it Baseline} stops assimilating new data of the emergence and flux transport takes over leading the region into cancellation and decay, while the two other simulations with far-side information continue adding significant flux and providing a more comprehensive description of the evolution of that region in the far-side.
  \item[12035] This is another active region with multiple far-side emergences in different rotations that are well represented by the {\it AFT+304} model. The evolution of this active region in the {\it AFT 304 only} reveals the challenge that a model without near side data assimilation has over long timescales. Active region 12035 emerges over an already large background flux in {\it AFT 304 only}. This is a consequence of a period of high activity in the belts. A nearby decaying active region rotates into the field of view of analysis, causing it to appear as if there is additional growth before the decay phase begins. 
\end{description}

We outlined in Section~\ref{sec:sims} that for each of the two sets of simulations we applied four different threshold caps for the amount of flux that is allowed to emerge per active region per 8-hour interval between STEREO/AIA 304 maps. The purpose of these caps is to limit the effect that large increases in intensity originating from large flares can have in the magnetic flux proxy. Figure~\ref{fig:AR_compare_304only} shows a comparison of the {\it AFT 304 only} for the three largest thresholds ($5\times10^{20}$, $1\times10^{21}$, $5\times10^{21}$ Mx). We find that a too stringent cap does not allow the model to capture the emergence of flux sufficiently rapidly, therefore underestimating the maximum total unsigned flux in the region. It is for that reason and for the sake of clarity that we do not show the $1\times10^{20}$ Mx case in Figure~\ref{fig:AR_compare_304only}. Our analysis indicates that a cap of $5\times10^{21}$ Mx strikes the best balance between limiting the effects of large flares while at the same time accurately capturing the emergence timescales in the active regions' flux evolution. This amounts to a flux emergence rate of about $6.25\times10^{20}$ Mx/hr, which would be sufficient to capture the largest emergence rates in \citet{2017norton} or the $2.5\times10^{20}$ Mx/hr upper limit we measured for active regions in \citet{ugarte-urra2012}. %We note that this threshold does not perform as well for the active region decay phase. More on this in the Discussion Section. {\color{red} Not sure about this conclusion. The far-side decay for AFT+304 seems reasonable. The problem is 304 only??.}
%Suggests that this amounts to a flux emergence rate (d_phi/d_t) of  $5\times10^{21}$ Mx per 8 hours which is  about  $6.25\times10^{20}$ Mx hr−1 , which is comparable to Hotta \& Iijima, 2020 ??? 

\subsection{Global Field} \label{subsec:global}
 
%SHOW results for three different simulations: Baseline, AFT+304, AFT 304 Only.
%Show Global Field properties: TUF, Flux Different, and Axial Dipole.

\begin{figure*}[ht] 
 \centering
 \includegraphics[width=0.95\textwidth,trim={40 100 40 75},clip]{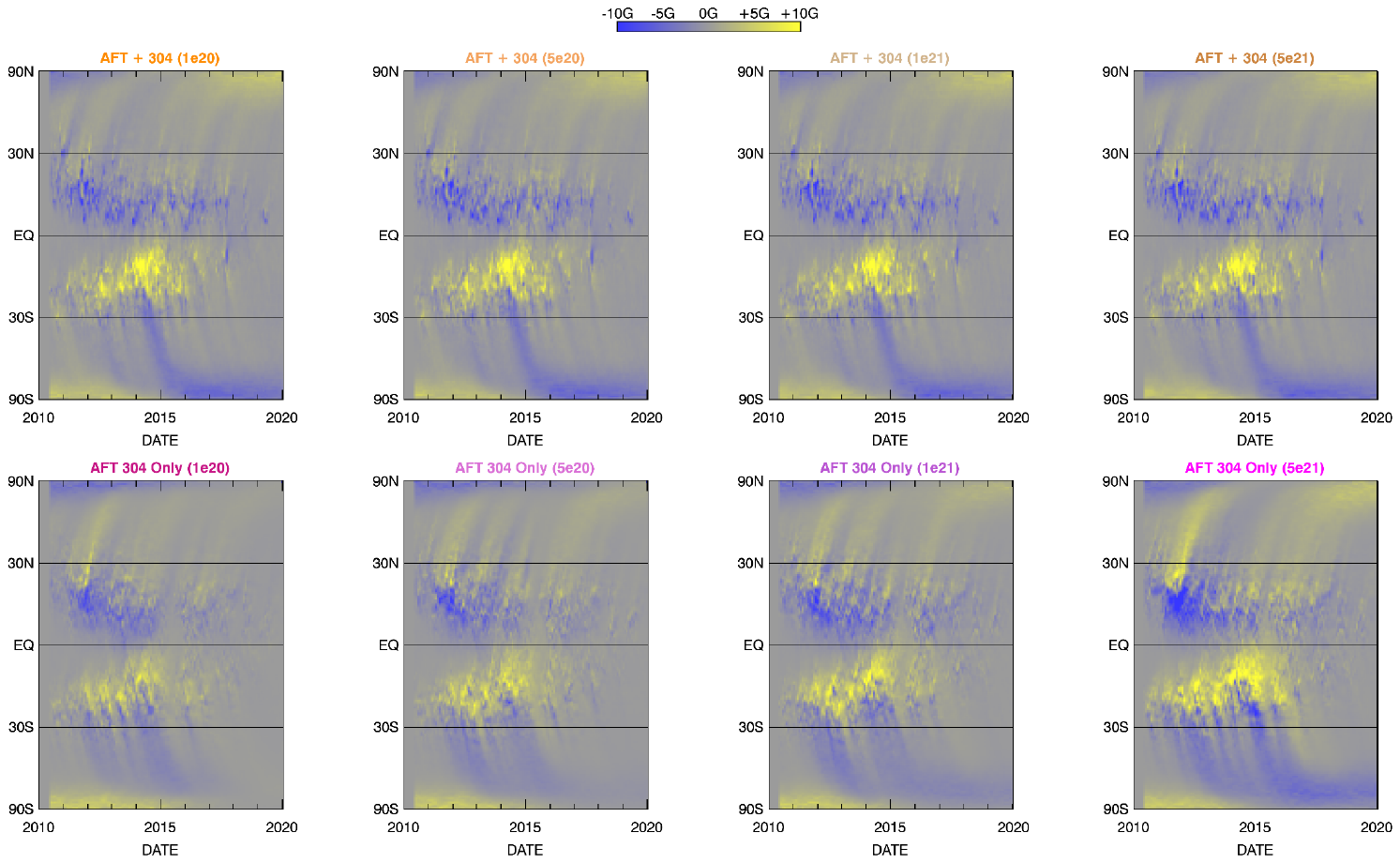}
 %\includegraphics[width=0.95\textwidth,trim={10 220 0 220},clip]{figures/flux_compare8_2022.pdf}
 %\caption{Magnetic Butterfly Diagrams.  NEED TO WRITE A CAPTION!!!!!}
 \caption{Magnetic Butterfly Diagrams . Magnetic butterfly diagrams for the {\it AFT+304} and {\it AFT 304 only} simulations are shown in the top and bottom rows respectively. \add[]{The titles of each panel are color coded based on} the flux cap (1e20, 5e20, 1e21, and 5e21) used.}
 \label{fig:compare1}
\end{figure*}

The active region evolution illustrates the flux evolution on small temporal and spatial scales, we now turn our attention to the flux evolution on large temporal and spatial scales. Here we examine the evolution of global solar properties over several years. We begin with a qualitative comparison by showing a magnetic butterfly diagram for eight of our simulations (shown in the top two rows of Figure ~\ref{fig:compare1}). The magnetic butterfly diagram is calculated by averaging the magnetic field at the surface over all longitudes and for each latitude and each Carrington Rotation.

%[trim=left bottom right top, clip] or ,trim={0 0 {.5\width} 0},clip

For the {\it AFT+304} simulations (top row, Figure ~\ref{fig:compare1}), the butterfly diagrams are nearly identical, thought some minor differences can be seen with careful inspection. There are some slight variations in field strength, but structurally they are essentially the same (e.g., the morphology of the dappled pattern in the butterfly wings, or the oscillating leading and following polarity polar streams). The striking similarity in these cases is due to the fact that the data assimilation process continually corrects the near-side of the simulation with the observations.  The butterfly diagram for the {\it Baseline} simulation is also nearly identical to these three and so is not shown. Any differences in the strength are due to the additional flux that is present on the far-side of the simulation, which is now included in the average for each latitude and each Carrington Rotation.

\begin{figure*}[ht] 
 \centering
 \includegraphics[width=0.95\textwidth,trim={22 95 08 80},clip]{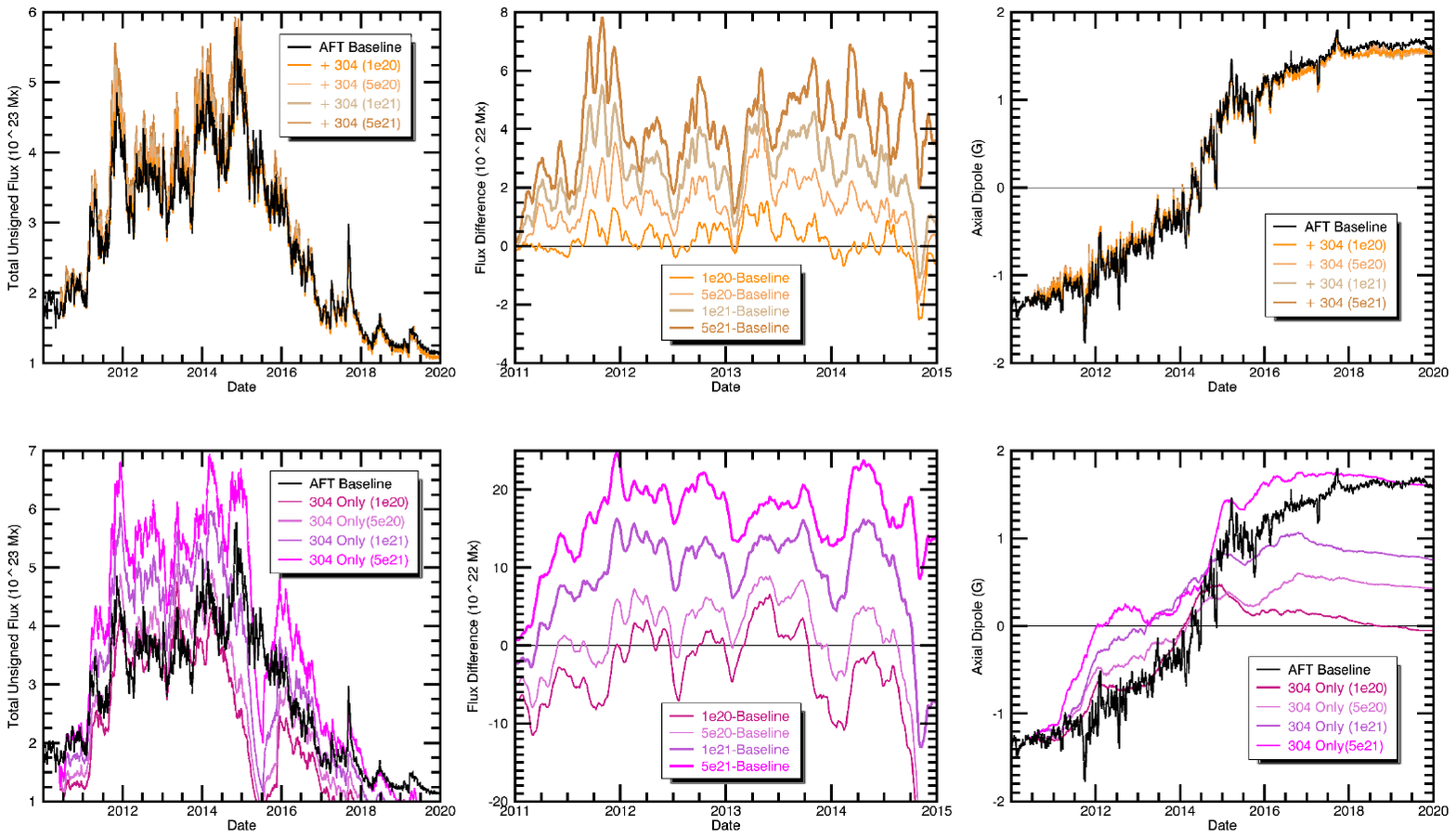}
 %\caption{Magnetic Butterfly Diagrams.  NEED TO WRITE A CAPTION!!!!!}
 \caption{Global Simulation Properties. Plots of the Total Unsigned Flux (left), Flux difference (relative to the {\it Baseline}, middle), and the Axial Dipole (right) are shown in the top ({\it AFT+304} simulations) and bottom ({\it AFT 304 only} simulations) rows. The plots are color coded based on the flux threshold (1e20, 5e20, 1e21, and 5e21) used. For reference, the measured {\it Baseline} TUF and Axial Dipole values are shown in black }
 \label{fig:compare2}
\end{figure*}

Differences in the magnetic butterfly diagram are more pronounced for the {\it AFT 304 only} cases (bottom row, Figure ~\ref{fig:compare2}) because there is no data assimilation process to regulate the simulations. The butterfly diagrams for the for {\it AFT 304 only} simulations have clear differences both in the amplitude of the field strength and also structurally. As the active region flux cap increases, the strength of the magnetic field in the butterfly wings saturate and the stronger flux concentrations become wider. This causes the wings of the butterfly to widen as well. As a consequence, more flux cancellation across the equator occurs and more following polarity flux fills the polar streams. This causes the polar fields to reverse more quickly and become stronger by the end of the cycle. 

Comparing the {\it AFT 304 only} cases to the {\it Baseline} or {\it AFT+304} simulations, we find that the {\it AFT 304 only} simulations have distinct differences in both the mottled pattern and in the polar streams. This is largely due to the fact that the Active Regions in the {\it AFT 304 only} case are all given the average Joy's Law tilt. Another difference is an apparent gap in the emergence of new flux for the {\it AFT 304 only} cases in 2015. This was caused by the previously mentioned superior solar conjunction. Afterwards, the flux emergence does pick up again somewhat, but remains notably weak for the remainder of the cycle. This is caused by the diminishing latitudinal coverage due to both the loss of the STEREO B spacecraft observations as well as the return of the STEREO A spacecraft to the near-side of the Sun, as illustrated in Figure ~\ref{fig:datagaps}. 

Despite the differences, some structural similarities persist. For instance, the general outline of the butterfly wings and the latitudes of the polarity inversion lines are very similar. Additionally, a few of the following polarity streams are present in all of the simulations (e.g., in the northern hemisphere from 2011-2013 and in the southern hemisphere from 2014-2016). Overall, the {\it AFT 304 only} are able to reproduce the butterfly diagram reasonable well considering no direct magnetic date (other than the initial map) were used, particularly for the first 4-5 years when nearly full coverage was available. 

Next we quantify the differences between the AFT simulations. We begin by calculating the Total Unsigned Flux (TUF) over the entire Sun, both near-side and far-side. The TUF for the {\it AFT+304} simulations (using four different flux caps) are shown by the orange/brown lines in the top left panel of Figure ~\ref{fig:compare2}. For reference, the TUF measured in the {\it Baseline} is shown in black. The TUF for the {\it AFT+304} simulations seems to be fairly consistent (i.e., matching the overall evolution) with the {\it AFT Baseline}) and across all four simulations. In each case, the higher flux cap increases the TUF by a fraction of the {\it Baseline} values, with the biggest increase (about $10\%$) seen during solar maximum. In order to better visualize this, we also plot the difference between the TUF in each simulation and the {\it AFT Baseline}) TUF (middle panels) from 2011-2015. The flux difference calculated by subtracting the TUF from the {\it AFT Baseline} from each of the the {\it AFT+304} simulations (top middle panel) represents an estimate of how much flux is missing from the {\it AFT Baseline} due to new far-side active region growth and emergence. The analysis of the active region evolution in the previous section showed that flux cap of $5\times10^{21}$ Mx strikes the best balance between capturing the active region emergence while limiting the effects of large flares. For this time period, the missing far-side flux is typically on the order of $4-6\times10^{22}$ Mx, or about the size of a fairly large active region. This is an estimate during the time leading up to and including Solar Cycle 24 maximum (April 2014), and it should be noted that Cycle 24 was considered a very small cycle (the smallest observed in 100 years). Larger cycles would be expected to have even more far-side emergence and this should be considered a lower estimate of the amount of flux missing from SFT models during solar maximum when only near-side data is included.

%Interestingly, late in the cycle (from 2018-2020) the {\it AFT+304} simulations have a little bit less flux than the {\it AFT Baseline}. This suggests that the inclusion of the far-side active regions during that time period are producing additional flux cancellation, potentially due to the presence of rogue active regions on the near side???

%[trim=left bottom right top, clip] or ,trim={0 0 {.5\width} 0},clip

Similarly, the TUF for the {\it AFT 304 only} simulations is shown by the pink/purple lines in the bottom left panel of Figure ~\ref{fig:compare2}. We note that flux emergence is underestimated prior to 2012 and again after the 2nd half of 2014 for all but the strongest flux cap used ($5\times10^{21}$ Mx). This is due to the lack of 304 coverage as the STEREO spacecrafts progress in their orbit (leading to 2012, and after 2015) and during the superior solar conjunction (2014-2015). During the superior solar conjunction in 2014-2015, the TUF drops well below the {\it AFT Baseline} in all four cases. In the later half of 2015, as the STEREO A spacecraft emerges from the conjunction, the largest threshold cases again significantly exceed the {\it AFT Baseline}, while the mid threshold cases ($5\times10^{20}$ Mx and $1\times10^{21}$ Mx) reach similar values and the $1\times10^{2}$ Mx threshold case falls well below the {\it AFT Baseline} TUF. In 2017, the TUF in all these cases begin a progressive decline below the {\it AFT Baseline} as STEREO A progresses towards the Earth vantage and far-side coverage is lost. 

The flux difference calculated by subtracting the TUF from the {\it AFT Baseline} from the {\it AFT 304 only} simulations (bottom middle panel) far exceeds the estimate obtained from the {\it AFT+304} simulations in the $1\times10^{21}$ Mx and $5\times10^{21}$ Mx threshold cases. On global time scales, these thresholds are not consistent with the Sun. The amplitude of the $1\times10^{20}$ Mx and $5\times10^{20}$ Mx threshold cases are more inline with the {\it AFT Baseline} and the {\it AFT 304 only} results, respectively. However, the structure in the TUF evolution is not well captured. This is due to a steady accumulation of flux from flaring regions that are not being corrected by the data assimilation of magnetograms.

Another quantitative metric to consider is the progression of the Sun's axial dipole moment. This quantity measures the polar field evolution - i.e., the decay, reversal, and build up of magnetic flux in the polar regions. The ability to reproduce the evolution of the polar fields is essential for SFT models, as the polar field at the time of solar cycle minimum is a key predictor of the strength of the next cycle \citep{1987SchattenSabatino, 2015Hathaway, upton2014a}. The axial dipole strength is given as

\begin{equation}
D(t) = (3/4\pi) \int B_r(\theta, \phi, t)  \: cos \theta \: d\Omega,
\end{equation}

where $\, B_r(\theta, \phi, t) \,$ is the radial photospheric magnetic field at latitudes $theta$, longitudes $\phi$, and time $t$ integrated over the solid angle $\Omega$ \citep{WangSheeley1991}. The axial dipole moment from each of the simulations is shown in the right column of Figure ~\ref{fig:compare2}, with the {\it AFT+304} simulations shown in the top right panel and the {\it AFT 304 only} simulations shown in the bottom right panel. In both cases, the axial dipole moment from the {\it AFT Baseline} is shown in black.

The axial dipole moment from all of the {\it AFT+304} simulations are nearly identical to the {\it AFT Baseline}, once again showing that the data assimilation process is able to correct for inaccuracies in the far-side emergence. For the {\it AFT 304 only} simulations, again, the $1\times10^{20}$ Mx and $5\times10^{20}$ Mx threshold cases do the best at reproducing the axial dipole evolution until 2015. During this same time period, the $1\times10^{21}$ Mx, and $5\times10^{21}$ Mx threshold cases have too much flux reaching the pole, causing the axial dipole to reverse too quickly. During 2014, following the superior solar conjunction, the dipole evolution falls briefly in all cases. While this is followed by a short period of growth, they all begin a steady decline in 2016 . This occurs because the loss of \ion{He}{2} 304\,\AA\ coverage results in an apparent gap in active region emergence, which allows leading polarity flux to be carried to the pole (see Figure ~\ref{fig:compare2}, bottom row) during the conjunction. Afterwards, active region emergence picks up briefly before the combined effects of the declining cycle and declining far-side coverage result in only a small amount of new flux is being added, primarily close to the equator. Since most of the flux cancellation occurs at the lowest latitudes, very little new flux reaches the poles. The flux that is transported at this stage is primary leading polarity flux, causing the weak decline in the polar fields.

Determining the best threshold case for the global parameters in the {\it AFT 304 only} simulations is not straight forward. The  $1\times10^{21}$ Mx threshold case is most consistent with the TUF in the best case from the {\it AFT+304} results, but does not do well at reproducing the polar field evolution. The $1\times10^{20}$ Mx case performs remarkably well at reproducing the polar field evolution, but underestimates the TUF. Choosing the best case for the {\it AFT 304 only} simulations is thus dependent on the end-use application. 

%Equatorial dipole strength is given as
%\begin{subequations}\label{first:main}
%\begin{equation}
%H(t) = [ \: (h_1(t))^2 + (h_2(t))^2 \:]^{1/2}, \tag{\ref{first:main}}
%\end{equation}
%where
%\begin{align}
%        h_1(t) = (3/4\pi) \int B_r(\theta, \phi, t)  \: sin \theta \: cos \phi \: d\Omega, \label{first:a} \\
%        h_2(t) = (3/4\pi) \int B_r(\theta, \phi, t)  \: cos \theta  \:  cos \phi  \: d\Omega . \label{first:b}
%\end{align}
%\end{subequations}

\section{Discussion}

\subsection{AR Detection} \label{subsec:ARfinder}

We have developed an automated detection algorithm to identify active regions in \ion{He}{2} 304\,\AA\ intensity images and incorporate them into the AFT SFT model. For this purpose, we created higher order (Level 2) processed files which included several corrections, including despiking the intensity to mitigate against flaring events. We found that the power-law relationship is sensitive to how the boundary of the active region is defined and to the magnetic flux density thresholds and have computed a new empirical relationship tailored to the {\it AFT Baseline} maps. Our automatic detection algorithm identifies the location and infers flux in active regions from the \ion{He}{2} 304\,\AA\ intensity images using this relationship. The algorithm them compares the active regions to the AFT maps to determine if new active region growth has occurred. When new growth occurs, idealized bipoles are added to the AFT maps. 

Despite  despiking the Level 2 \ion{He}{2} 304\,\AA\ intensity images, we found that flaring events still occurred frequently. To further mitigate against this, we implemented a cap on how much a given active region was allowed to grow in the 8 hour interval between Level 2 \ion{He}{2} 304\,\AA\ intensity images. We then conduced a series of simulations to investigate the performance of the automated detection algorithm with several flux cap thresholds: $1\times10^{20}$ Mx, $5\times10^{20}$ Mx, $1\times10^{21}$ Mx, and 5$\times10^{21}$ Mx. For each of these flux caps, we ran AFT with and without data assimilation,  {\it AFT+304} and {\it AFT 304 only} respectively. The latter of which is informed exclusively by the \ion{He}{2} 304\,\AA\ intensity maps.

\subsection{AFT Simulations} \label{subsec:AFTSims}

Results from the {\it AFT+304} and {\it AFT 304 only} simulations showed that the $5\times10^{21}$ Mx threshold case (corresponding to as active region growth rate of about $6.25\times10^{20}$ Mx/hr) performed the best at capturing the Active Region growth. 
%\textbf{\color{red}[Should this sentence also include AFT+304? The follow-up discussion in this paragraph correctly applies to 304 only, but I wonder if it should start first stressing that AFT+304 performs well with the $5\times10^{21}$ cap before moving to the 304 only problems]}. 
However, the global parameters in the {\it AFT 304 only} simulations did not fare as well: the TUF was much too high and the axial dipole reversed polarity much too quickly, suggesting too much flux is being transported to the poles. Figure~\ref{fig:AR_compare_304only} shows that while this threshold does well at capturing the active region growth phase, the active region decay phase is not as well reproduced. This can be attributed to a few different factors, including residual flaring events, active region properties, and small scale flux interactions. Flaring events will cause excess flux to be deposited into the active region, directly resulting in more residual flux to be transported poleward. The active regions in these simulations are incorporated into AFT as idealized Gaussian spots with tilts and separation distances determined from statistical properties. Differences between the simulated and observed tilts and separations will change the axial dipole moment of the active regions, which could also result in more flux being transported to the poles. Another possibility is that interactions with nearby small scale flux, not included in the AFT idealized active regions, may play a non-trivial role in the flux cancellation process. \citet{Yeates2020} showed that idealized bipoles, as opposed to observed complex morphologies, could also lead to an overestimation of the axial dipole.   \citet{ugarte-urra2015} found that AFT was able to reproduce the evolution of the active regions to within a factor of 2 when starting with the observed active region morphology. We find that the active regions in the {\it AFT+304} performed even better (see Figure~\ref{fig:AR_compare_304only}), suggesting that the idealized active region are the most likely source of this discrepancy. %\textbf{\color{red}[Fig.5 suggests that with this new model we do better than the factor 2 of the previous paper (HMI not shown though). I would not insist on that result. Remove? I think the difference between the idealized non-assimilated 304-ony vs real data is the likely culprit]}. 
\textbf{These results demonstrates that while \ion{He}{2} 304\,\AA\ does remarkably well at driving the SFT model, these data alone are insufficient.}
%the {\it AFT 304 only} simulations highlighted the need to find a balance between sufficient AR growth rate and mitigating flaring effects that cause excess flux in the system, the {\it AFT+304}

We find that the {\it AFT+304} simulations fared much better. This configuration provides the most accurate representation of the flux distribution on the entire surface of the Sun by combining the benefits of incorporate far-side emergence while using data assimilation of magnetograms to correct for inconsistencies in the implementation of the far-side Active Regions. Using the $5\times10^{21}$ Mx threshold case, \textbf{we estimate that for Cycle 24 surface flux transport models are missing approximately $4-6\times10^{22}$ Mx from far-side Active Regions.} While not implemented here, we intend to use these maps as the inner boundary condition for various types of atmospheric and coronal modeling in order to determine the impact of including this missing flux in making space weather predictions. \textbf{\ion{He}{2} 304\,\AA\ images can be used as a supplemental proxy for magnetic flux measurements when direct magnetic field observations are not complete.}

\subsection{Future Work} \label{subsec:Future}

While STEREO is currently positioned close to the sun-earth line, which limits the utility of the results for new applications, in a few years STEREO A will begin returning far-side images. Studies that use the older STEREO data, such as this one, are essential for determining how we can best use the far-side EUV observations when they become available again. In addition, these methodologies and results can be used to validate \add[]{other far-side proxies of the far-side, such as those obtained by helioseismic and Machine Learning/Artificial Intelligence (ML/AI) methodologies, as well as adapted for implementation with far-side observations from newer missions.}

\add[]{While helioseismic results show considerable promise, they are less reliable than direct imaging. Direct imaging in the EUV can precisely confirm the existence, location, and relative size of active regions. Helioseismic results, on the contrary, are still plagued with false positives and false negatives. Even far-side active region detections that are confirmed by direct far-side imaging or subsequent rotation on the near-side, possess large uncertainties in relevant parameters such as active region size, strength, location. The EUV automated far-side detection process described here can be used to create reliable active region catalogs. These catalogs can then be used to assess the uncertainty in far-side helioseismology active region detections and aid in making decisions on how best to include far-side results in SFT models.

The advent of ML/AI is providing new avenues for advancing helioseismic techniques and improving the accuracy. Both HMI and GONG helioseismology teams and we are currently exploring ML/AI methodologies to provide improved data products, including helioseismic inferred far-side magnetic maps} \citep{Chen_etal2022, Creelman_etal2024}. \add[]{We are currently working with both of these teams to adapt the techniques described in this paper for use with the helioseismic data, which we plan to address in follow-on papers. In addition to the advances ML/AI techniques offer to heliosesmology, these techniques are also being used to improve other existing data products. For example, ML/AI groups training on combined AIA/STEREO EUV images to create AI-generated magnetograms are consistent with HMI magnetograms} \citep{Kim_etal2019, Jeong_etal2022}. \add[]{Using our automated detection codes with these data to populate AFT with far-side active regions may prove more accurate than with the EUV observations directly. Another group is using ML/AI to cross-calibrate Hinode/SOT-SP data} \citep{Lites_etal2013} \add[]{with HMI data in order to create improved vector magnetograms}\citep{Fouhey_etal2023, Wang_etal2024} \add[]{. These new vector magnetograms are expected to minimize issues with the 180 dis-ambiguity as well as limb effects, problems which have long stood in the way of making vector date viable for assimilation into SFT models } 

\add[]{Finally, the methodologies developed here may be adapted for and improved by new direct far-side observations. For example, Solar Orbiter captures high-resolution EUV images as well as magnetograms. This data combination provide a unique opportunity to perform additional studies like this one. Coordinated observations can be used to cross-calibrate observations made by the two instruments and develop the ability to create proxies that can be used when one or the other is not observing the same target. These data, when available, could then be incorporated in SFT models to provide supplementary knowledge of far-side active region evolution and to access helioseismic results}\citep{Yang_etal2023} \add[]{.}

%% --- ACKNOWLEDGMENTS

\acknowledgments
We would like to thank the anonymous referee for valuable comments and suggestions that improved the quality of the paper.
AIA and HMI data are courtesy of NASA/SDO and the AIA and HMI science teams. %The SECCHI data are produced by an international consortium of the NRL, LMSAL and NASA GSFC (USA), RAL and U. Bham (UK), MPS (Germany), CSL (Belgium), IOTA and IAS (France). 
L.A.U.\ was supported by NASA Heliophysics Guest Investigator NNH15ZDA001N-HGI, NASA Heliophysics Living With a Star grants NNH16ZDA010N-LWS and NNH18ZDA001N-LWS, and by NASA grant NNH18ZDA001N-DRIVE to the COFFIES DRIVE Center managed by Stanford University. I.U.U. and H.P.W.\ were supported by the NASA Heliophysics Guest Investigator and Living With a Star programs and the Office of Naval Research. 
D.H.H.\ was supported by NASA contract NAS5-02139 (HMI) to Stanford University.

\vspace{5mm}
\facilities{SDO (AIA,HMI), STEREO (EUVI)}
\software{SolarSoft \citep{freeland2012}, VSO \citep{2009hill}}

% --- REFERENCES
\bibliography{main.bib}

\end{document}